\documentclass[aps,prl,twocolumn,superscriptaddress]{revtex4}
\usepackage{graphicx}
\usepackage{dcolumn}
\usepackage{bm}
\usepackage{verbatim}
\usepackage{color}
\usepackage[colorlinks]{hyperref}
\input{epsf}

\usepackage[sort&compress]{natbib}
\begin{document}


\title{Observation of linear-polarization-sensitivity in the microwave-radiation-induced magnetoresistance oscillations}

\author{R. G. Mani}
\author{A. N. Ramanayaka}
\affiliation{Department of Physics and Astronomy, Georgia State
University, Atlanta, GA 30303.}

\author{W. Wegscheider}
\affiliation{Laboratorium f\"{u}r Festk\"{o}rperphysik, ETH
Z\"{u}rich, 8093 Z\"{u}rich, Switzerland}

\date{\today}

\begin{abstract}
In the quasi two-dimensional GaAs/AlGaAs system, we investigate
the effect of rotating \textit{in-situ} the electric field of
linearly polarized microwaves relative to the current, on the
microwave-radiation-induced magneto-resistance oscillations. We
find that the frequency and the phase of the photo-excited
magneto-resistance oscillations are insensitive to the
polarization. On the other hand, the amplitudes of the
magnetoresistance oscillations are remarkably responsive to the
relative orientation between the microwave antenna and the
current-axis in the specimen. The results suggest a striking
linear-polarization-sensitivity in the radiation-induced
magnetoresistance oscillations.
\end{abstract}

\maketitle

\section{introduction}

High quality quasi two-dimensional electron systems (2DES)
realized in GaAs/AlGaAs semiconductor heterostructures have long
served to examine intriguing phenomena.\cite{grid-2, grid-1} The
rich new physics in this material system has animated further
improvements in material quality which, in turn, has driven new
developments within the field. Physical phenomena induced by
microwave and terahertz photo-excitation at high filling factors
or low magnetic fields, $B$, may be counted amongst these
developments. Here, the realization of radiation-induced
$B^{-1}$-periodic magnetoresistance oscillations and associated
zero-resistance states led to broad experimental \cite{ grid1,
grid2, grid3, grid4, grid5, grid6, grid7, grid8, grid9, grid11,
grid12, grid13, grid15, grid16, grid17, grid19, grid20, grid21,
grid22} and theoretical\cite{grid23, grid24, grid25, grid26,
grid101, grid27, grid28, grid29, grid30, grid31, grid32, grid33,
grid34, grid35, grid36, grid37, grid38, grid39, grid40, grid42,
grid43, grid44, grid45, grid46, grid47, grid48, grid49}
investigations of transport in the photo-excited 2DES.

The microwave and terahertz radiation-induced magneto-resistance
oscillations in the 2DES are characterized by $B^{-1}$ periodic
oscillations in the diagonal magnetoresistance, $R_{xx}$, of the
2DES at cryogenic temperatures, $T$. These $R_{xx}$ oscillations
show a strong sensitivity to $T$ and the microwave power, $P$, at
modest $P$. Proposed mechanisms for such oscillations include
radiation-assisted indirect inter-Landau-level scattering by
phonons and impurities (the displacement model),\cite{grid23,
grid25, grid27, grid46} non-parabolicity effects in an ac-driven
system (the non-parabolicity model),\cite{grid101} a
radiation-induced steady state non-equilibrium distribution (the
inelastic model),\cite{grid33} and the periodic motion of the
electron orbit centers under irradiation (the radiation driven
electron orbit model).\cite{grid34, grid37}

Under typical experimental conditions, some or all of these
mechanisms can contribute towards sufficiently large amplitude
radiation-induced magneto-resistivity oscillations such that,
theoretically, at the oscillatory minima, the magneto-resistivity
is able to take on negative values. According to theory, negative
resistivity triggers, however, an instability in the uniform
current distribution, leading to current domain formation, and the
experimentally observed zero-resistance states.\cite{grid24,
grid43}

Although these theories suggest radiation-induced
magneto-resistance oscillations, they differ with respect to their
predictions on, for example, the microwave polarization
sensitivity in the radiation-induced oscillations. Here, the
displacement model predicts that the oscillation-amplitude depends
on whether the microwave electric field, $E_{\omega}$, is parallel
or perpendicular to the $dc$-electric field,
$E_{DC}$.\cite{grid25} On the other hand, the inelastic model
unequivocally asserts polarization insensitivity to the
radiation-induced magneto-resistance oscillations.\cite{grid33}
Polarization immunity, in the radiation-driven electron orbit
model, depends parametrically upon the damping factor, $\gamma$,
exceeding the frequency of the microwave field.\cite{grid37}
Finally, the non-parabolicity model suggests a perceptible
polarization sensitivity for linearly polarized
microwaves,\cite{grid101} while indicating the absence of such
oscillations for circularly polarized radiation.\cite{grid101}
From the experimental perspective, previous work on L-shaped Hall
bars indicated that the frequency and phase of the
radiation-induced magnetoresistance oscillations are insensitive
to the microwave polarization.\cite{grid5} Other work on
square-shaped specimens asserted the insensitivity of the
microwave induced magneto-resistance oscillations to the
polarization sense of circularly and linearly polarized
microwaves.\cite{grid13}

\begin{figure}[t]
\begin{center}
\leavevmode \epsfxsize=3.25 in \epsfbox {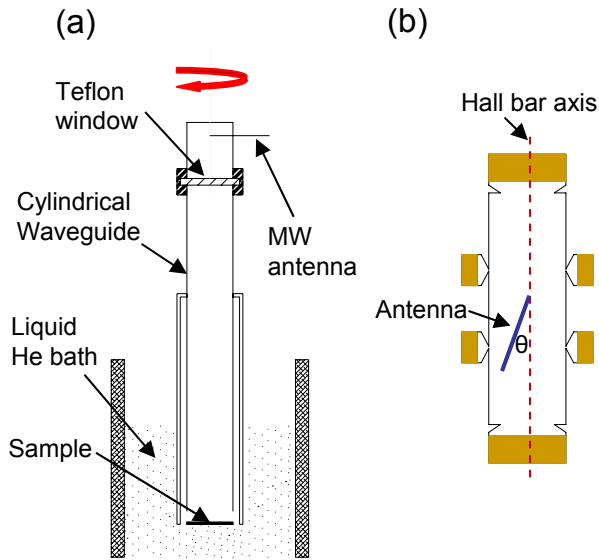}
\end{center}
\caption{ The experimental setup: (a) A microwave launcher, which
includes a monopole-probe-coupled microwave (MW) antenna, is free
to rotate about the axis of a cylindrical waveguide. A flexible
semi-rigid coax (not shown) couples the MW-antenna to the
microwave source. The teflon window serves to isolate the
evacuated section of the sample holder. (b) A Hall bar specimen,
shown as "sample" in (a), is installed such that the Hall bar axis
is parallel to the MW-antenna for $\theta = 0^{0}$.\label{afig:
epsart}}
\end{figure}

Here, we investigate the effect of rotating, \textit{in-situ}, the
polarization of linearly polarized microwaves relative to
long-axis of Hall bars. Strikingly, we find that the amplitude of
the radiation-induced magneto-resistance oscillations are
remarkably responsive to the relative orientation between the
linearly polarized microwave electric field and the current-axis
in the specimen. The results appear qualitatively consistent with
the displacement, the non-parabolicity, and the radiation driven
electron orbit model for $\gamma < \omega$.

\begin{figure}[t]
\begin{center}
\leavevmode \epsfxsize=3.25 in \epsfbox {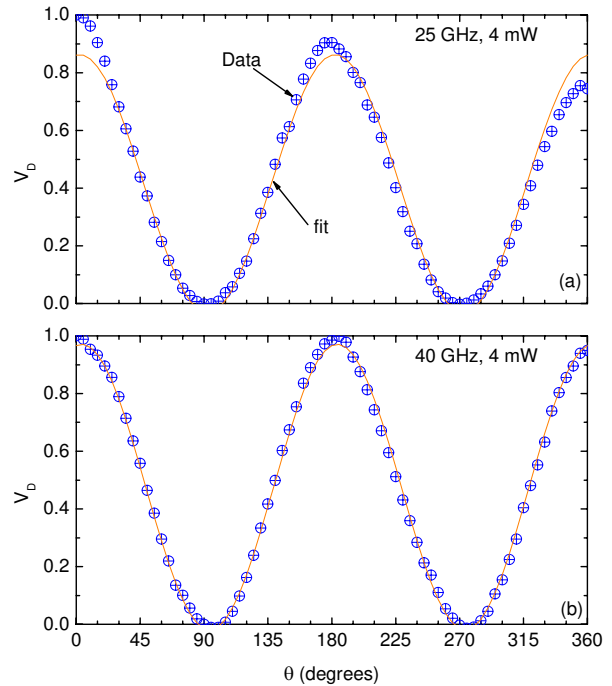}
\end{center}
\caption{ Figs. 2(a) and 2(b) show the normalized response
($V_{D}$) of the diode detector (circles) placed at the sample
position,  at $f = 25 GHz$ and $f = 40 GHz$, respectively. For
these measurements, the sample in Fig. 1(a) was replaced with an
analyzer consisting of a monopole-probe-coupled-antenna and a
square-law diode-detector assembly. Thus, Fig.2(a) and 2(b) show
that the microwave polarization is preserved from the MW antenna
to the sample position.\label{fig: epsart}}
\end{figure}

\section{Experiment and Results}

Experimental microwave polarization studies in this context are
difficult to carry out since it is non-trivial to rotate, in-situ,
a specimen with wires over $360^{0}$ at the end of a $2 m$ long
sample holder, within a small ($\approx 30mm$) diameter low
temperature cryostat. To overcome this barrier, we have developed,
instead, a setup where the wired sample remains fixed within the
cryostat, while the microwave polarization is rotated with respect
to the sample, from outside the cryostat. To achieve this
capability, see Fig. 1, the canonical rectangular waveguide was
replaced with a circular ($\approx 11mm$ i.d.) waveguide, and a
rotatable coax-to-waveguide-adapter, probe-coupled,
electric-monopole-antenna, microwave-launcher [MW-antenna in Fig.
1(a)] was developed to couple microwaves into the waveguide. Here,
the angular position of the MW-antenna could be set as desired and
then locked in place with a clamped quick connect. The Hall bar
sample was mounted at the low temperature ($T=1.5K$) end of the
circular waveguide as shown in Fig. 1(a), and the long axis of the
device was oriented parallel to polarization axis of the
MW-antenna. Thus, $\theta$, [see Fig. 1(b)], represents the
rotation angle of the MW-antenna with respect to the device
long-axis. These Hall bars, with a width $W =400\mu m$, were
characterized by n (4.2K) = 2.2$\times$$10^{11}$ $cm^{-2}$ and
$\mu \approx 8\times 10^{6}cm^{2}/Vs$.  The four-terminal diagonal
resistance, $R_{xx} = V_{xx}/I$, was extracted from $V_{xx}$
measurements between adjacent diagonal voltage contacts, [see Fig.
1(b)], as the current $I$ was applied via the ends. Thus, the
length (L)-to-width (W) ratio for the $R_{xx}$ measurements was
$L/W = 1$, see Fig. 1.

\begin{figure}[t]
\begin{center}
\leavevmode \epsfxsize=3.25 in \epsfbox {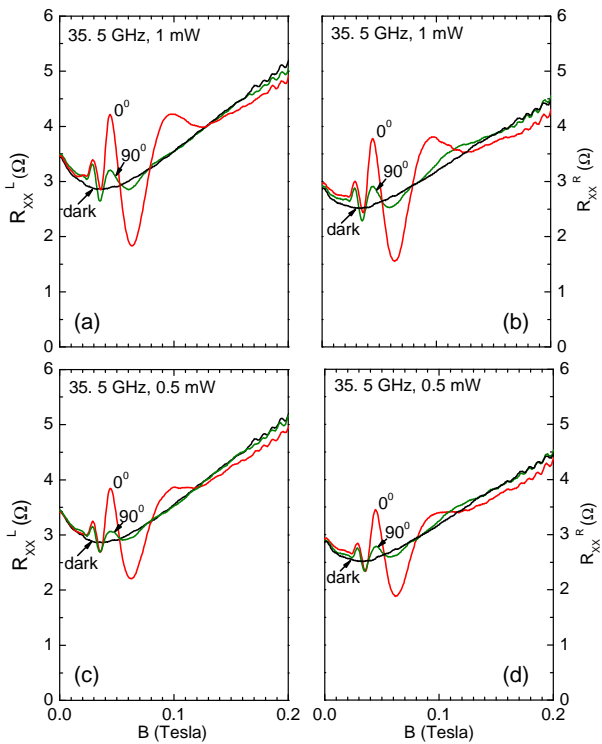}
\end{center}
\caption{ Microwave-induced magneto-resistance oscillations in
$R_{xx}$ at $1.5K$ are shown at $f = 35.5 GHz$ for $P = 1 mW$ in
panels (a) and (b)  and for $P = 0.5 mW$ in panels (c) and (d),
for sample-$1$. The $R_{xx}$ measured on the left (right) side of
the Hall bar, see Fig. 1, is shown as $R_{xx}^{L}$ ($R_{xx}^{R}$).
Each panel shows a set of three traces of $R_{xx}$ vs. $B$: a dark
curve (black), a curve (red) obtained at $\theta = 0^{0}$, and a
trace (green) obtained at $\theta = 90^{0}$. All panels exhibit
reduced amplitude radiation-induced magneto-resistance
oscillations at $\theta = 90^{0}$.\label{fig:epsart}}
\end{figure}

Some questions of interest here include whether polarized
microwaves are produced by the MW antenna, and whether this
polarization is preserved to the specimen. In order to answer
these questions, preliminary tests were carried out using an
"analyzer" consisting of an electric-monopole,
probe-coupled-antenna and square law detector. Bench tests carried
out with the MW-antenna [Fig. 1(a)] and the "analyzer" indicated
that polarized microwaves were generated by the microwave
launcher. For further tests, this "analyzer" was placed at the
sample end of the sample holder and fixed at a particular
orientation, as the MW-antenna was rotated through $360^{0}$ at
$5^{0}$ increments. Fig. 2(a) and 2(b) show the normalized
detector response, $V_{D}$, of the diode detector at $f = 25 GHz$
and $f = 40 GHz$. The figure exhibits the expected sinusoidal
variation, i.e., $V_{D} \propto cos^{2}{\theta}$, for linearly
polarized radiation, of the received power as a function $\theta$.
Also shown in Fig. 2 are fits to $V_{D} = A + Bcos^2(\theta +
\theta_{0})$. We find, for Fig. 2(a), $A = 0.0$, $B = 0.9$, and
$\theta_{0}= -2.4^{0}$, and for Fig. 2(b), $A = 0.0$, $B = 1.0$,
and $\theta_{0} = -4.0^{0}$. Here, $\theta_{0}$ is within
experimental uncertainties. Thus, polarized radiation is generated
at the launcher and the polarization is preserved down to the
sample.
\begin{figure}[t]
\begin{center}
\leavevmode \epsfxsize=3.25 in \epsfbox {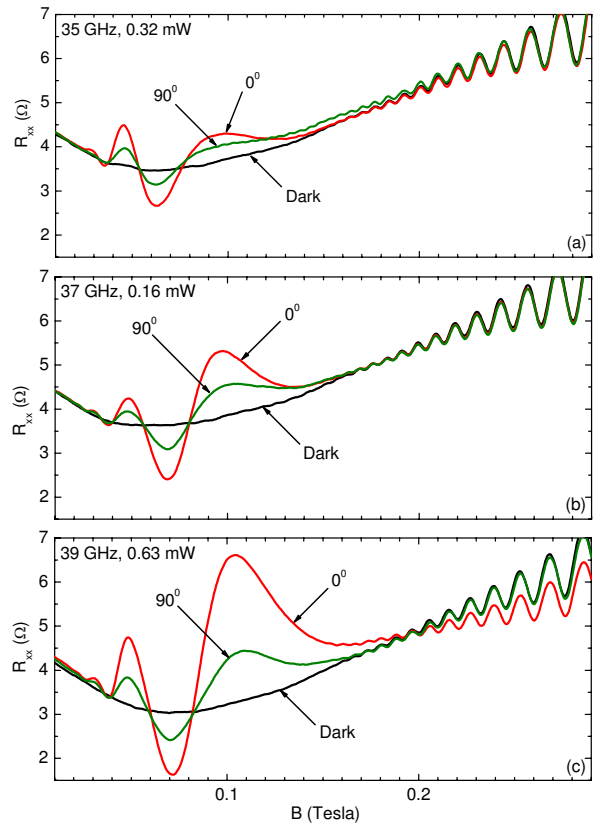}
\end{center}
\caption{ Microwave induced magneto-resistance oscillations in
$R_{xx}$ are shown at (a) $f = 35 GHz$, (b) $37 GHz$, and (c) $39
GHz$ at $1.5K$ for sample-$2$. Each panel shows a set of three
traces: a dark curve (black), a curve (red) obtained at $\theta =
0^{0}$, and a trace (green) obtained at $\theta =
90^{0}$.\label{fig:epsart}}
\end{figure}

Figure 3 exhibits the $R_{xx}$ vs. $B$ at $f = 35.5 GHz$, with the
Hall bar sample ($sample-1$) in place at the bottom of the
waveguide sample holder. Figures 3(a) and (b) show the results
obtained at a source-power $P = 1 mW$, while Figs. 3(c) and(d)
show the same obtained at $P = 0.5 mW$. Here, $R_{xx}^{L}$ and
$R_{xx}^{R}$ represent the measurement on the left (L) and right
(R) sides of the device(Fig. 1). Each panel of Fig. 3 includes
three traces: A dark trace (in black) obtained in the absence of
microwave photoexcitation. A $\theta = 0^{0}$ trace in red, where
the MW antenna is parallel to the long-axis of the Hall bar. A
comparison of Fig. 3(a) and (c) [or Fig. 3(b) and (d)] shows that
the $0^{0}$ (red) traces exhibit larger-amplitude
radiation-induced magneto-resistance oscillations at $P = 1 mW$
than at $P = 0.5 mW$. This feature corresponds to the usual
observation that the oscillation-amplitude increases with $P$ at
modest photo-excitation.\cite{grid21, grid49} Finally, the panels
of Fig. 3 also exhibit, in green, the $\theta = 90^{0}$ traces,
where the MW antenna is perpendicular to the long-axis of the Hall
bar. Again, as expected, a comparison of Fig. 3(a) and (c) [or
Fig. 3(b) and (d)] shows that the $90^{0}$ traces exhibit larger
amplitude radiation-induced magneto-resistance oscillations at $P
= 1 mW$ [Fig. 3(a)] than at $P = 0.5 mW$ [Fig. 3(b)].

The remarkable feature is observed when one compares the red
($0^{0}$) and green ($90^{0}$) traces within any single panel of
Fig. 3. Such a comparison indicates that the amplitude of the
radiation-induced magneto-resistance oscillations is reduced at
the $\theta = 90^{0}$ MW antenna  orientation. Thus far, our
experiments have shown upto a factor-of-ten reduction in the
oscillation amplitude under polarization rotation. Although the
magneto-resistance oscillations are reduced in amplitude,
typically, they are not completely extinguished at $\theta =
90^{0}$.  Finally, the period and the phase of the
radiation-induced magnetoresistance oscillations are unchanged by
MW-antenna rotation; this feature is readily apparent in Fig. 3.
Figure 4 illustrates similar measurements in a second specimen
($sample-2$) at $35 GHz$ [Fig. 4(a)], $37 GHz$ [Fig. 4(b)], and
$39 GHz$ [Fig. 4(c)] at various $P$. As is evident in Fig. 4, the
oscillation amplitude is again reduced for $\theta = 90^{0}$.

\section{Discussion}

In considering the implications, it is necessary to note that,
since $\omega \tau >> 1$ over the range of $B$ where the
photo-excited resistance oscillations are observed, the
$dc$-electric field should be oriented nearly perpendicular to the
Hall bar axis (see Fig.1). In the displacement model of
ref.\cite{grid25}, the inter-Landau level contribution to the
photo-current includes a term with a Bessel function whose
argument depended upon whether $E_{DC}$ and $E_{\omega}$ are
parallel or perpendicular to each other. Hence, the dissipative
microwave photoconductivity can exhibit polarization selectivity
in the displacement model. According to ref. \cite{grid33}, for
$\tau_{in} >> \tau_{q}$, where $\tau_{in}$ and $\tau_{q}$ are the
inelastic- and the single particle- relaxation times,
respectively, a larger contribution to the amplitude of the
radiation-induced magneto-oscillations is provided by the
inelastic mechanism than by the displacement mechanism. Further,
Ref. \cite{grid33} indicated that the inelastic mechanism does not
depend on the orientation of the linear polarization of the
microwave field, unlike the displacement mechanism, which also
yields a $T$-independent contribution to the oscillatory
conductivity. In their radiation driven electron orbit
model,\cite{grid37} polarization immunity is realized when $\gamma
> \omega = 2\pi f$. Finally, the non-parabolicity model included a
strong polarization sensitivity, with the dissipation an odd
function of the de-tuning from cyclotron resonance.\cite{grid101}

Based on the above, these experimental results appear
qualitatively similar to expectations based on a "displacement" or
"non-parabolic" or a "radiation-driven electron orbit" term with
$\gamma < \omega$. Yet, the experimental feature that the
oscillations do not vanish completely at $\theta = 90^{0}$ seems
not to rule out, at least at this stage of experimentation, the
existence of a linear-polarization-immune-term in the
radiation-induced transport. Next, we address the report of linear
polarization immunity in Ref. \cite{grid13}. Those measurements
were apparently carried out on $4\times4 mm^{2}$ square shaped
specimens, with a length to width ratio of one.\cite{grid13} In
such a square specimen with point contacts, the current stream
lines are expected to point in different directions over the face
of the sample. Then, the variable angle between the linear
microwave polarization and the local current orientation could
possibly serve to produce an effectively polarization averaged
measurement.

Finally, we comment upon expectation for the relative
sensitivity/immunity of the radiation-induced magneto-resistance
oscillations to the sense of circular polarization in the Hall bar
geometry where the current orientation is presumably well defined,
given the remarkable sensitivity to linear polarization shown
here: Recall that both senses of circularly polarized radiation
can be decomposed, into one linearly polarized wave that is
polarized parallel to the long axis, and another $90^{0}$ phase
shifted linearly polarized wave that is polarized parallel to the
short axis of the device. Since the linearly polarized component,
which is responsible for stimulating the radiation-induced
magnetoresistance oscillations, occurs in both decompositions,
immunity of the radiation-induced magneto-resistance oscillations
to the sense of circular polarization seems plausible even when
there is a strong sensitivity to the sense of linear polarization
in the Hall bar.

\section{Acknowledgements}

This work has been supported by A. Schwartz and the DOE under
DE-SC0001762 and by D. Woolard and the ARO under W911NF-07-01-015.
Thanks to T. Ghanem for the help with the experimental
development.


\begin{thebibliography}{53}
\bibitem{grid-2} R. E. Prange and S. M. Girvin, The Quantum Hall
Effect, 2nd. ed. (Springer, New York, 1990).
\bibitem{grid-1} S. Das Sarma and A. Pinczuk, Perspectives in
Quantum Hall Effects (Wiley, New York, 1996).

\bibitem{grid1} R. G. Mani, J. H. Smet, K. von Klitzing, V. Narayanamurti, W.
B. Johnson, and V. Umansky, Nature (London)  420, 646 (2002).

\bibitem{grid2} M. A. Zudov, R. R. Du, L. N. Pfeiffer, and K. W. West, Phys.
Rev. Lett. 90, 046807 (2003).

\bibitem{grid3}R. G. Mani, V. Narayanamurti, K. von Klitzing, J. H. Smet, W. B. Johnson, and V. Umansky, Phys. Rev. B 69, 161306 (2004); 70, 155310
(2004).

\bibitem{grid4} R. G. Mani et al., Phys. Rev. Lett. 92, 146801 (2004);
Phys. Rev. B 69, 193304 (2004).

\bibitem{grid5}R. G. Mani, Physica E (Amsterdam) 22, 1 (2004); 25, 189
(2004); Appl. Phys. Lett. 85, 4962 (2004).

\bibitem{grid6}S. A. Studenikin et al., Sol. St. Comm. 129, 341 (2004).

\bibitem{grid7}A. E. Kovalev et al., Sol. St. Comm. 130, 379 (2004).

\bibitem{grid8}R. L. Willett et al., Phys. Rev. Lett. 93, 026804
(2004).

\bibitem{grid9} R. R. Du et al., Physica E (Amsterdam) 22, 7 (2004).

\bibitem{grid11} R. G. Mani, IEEE Trans. Nanotechnol. 4, 27 (2005);
Phys. Rev. B 72, 075327 (2005); Sol. St. Comm. 144, 409 (2007);
Appl. Phys. Lett. 92, 102107 (2008); Physica E 40, 1178
(2008).

\bibitem{grid12} B. Simovic et al., Phys. Rev. B 71, 233303
(2005).

\bibitem{grid13} J. H. Smet et al., Phys. Rev. Lett. 95, 116804
(2005).

\bibitem{grid15} Z. Q. Yuan et al., Phys. Rev. B 74, 075313 (2006).

\bibitem{grid16} R. G. Mani, Appl. Phys. Lett. 91, 132103 (2007).

\bibitem{grid17} S. A. Studenikin et al., Phys. Rev. B 76, 165321
(2007).

\bibitem{grid19} A. Wirthmann et al., Phys. Rev. B 76, 195315 (2007).

\bibitem{grid20} S. Wiedmann et al., Phys. Rev. B 78, 121301(R)
(2008).

\bibitem{grid21} R. G. Mani et al., Phys. Rev. B 81, 125320 (2010); ibid. 79, 205320
(2009); A. N. Ramanayaka, R. G. Mani, and W. Wegscheider,
\textit{ibid.} 83, 165303 (2011).

\bibitem{grid22} O. M. Fedorych et al., Phys. Rev. B 81, 201302 (2010).

\bibitem{grid23} A. C. Durst et al., Phys. Rev. Lett. 91, 086803 (2003).

\bibitem{grid24} A. V. Andreev et al., Phys. Rev. Lett.
91, 056803 (2003).

\bibitem{grid25} V. Ryzhii and R. Suris, J. Phys.: Cond. Matt. 15, 6855 (2003).

\bibitem{grid26} V. Ryzhii and A. Satou, J. Phys. Soc. Jpn. 72, 2718 (2003).

\bibitem{grid101} A. A. Koulakov and M. E. Raikh, Phys. Rev. B 68,
115324 (2003).

\bibitem{grid27} X. L. Lei and S. Y. Liu, Phys. Rev. Lett. 91, 226805 (2003).

\bibitem{grid28} P. H. Rivera and P. A. Schulz, Phys. Rev. B 70, 075314
(2004).

\bibitem{grid29} X. L. Lei, J. Phys.: Condens. Matter 16, 4045 (2004).

\bibitem{grid30} S. A. Mikhailov, Phys. Rev. B 70, 165311 (2004).

\bibitem{grid31} J. Inarrea and G. Platero, Phys. Rev. B 72, 193414 (2005).

\bibitem{grid32} X. L. Lei and S. Y. Liu, Phys. Rev. B 72, 075345 (2005).

\bibitem{grid33} I. A. Dmitriev et al., Phys. Rev. B 71, 115316 (2005).

\bibitem{grid34} J. Inarrea and G. Platero, Phys. Rev. Lett. 94, 016806 (2005).

\bibitem{grid35} A. Auerbach et al., Phys. Rev.
Lett. 94, 196801 (2005).

\bibitem{grid36} J. Inarrea and G. Platero, Appl. Phys. Lett. 89, 052109
(2006); ibid. 90, 172118 (2007); ibid. 92, 192113 (2008).

\bibitem{grid37} J. Inarrea and G. Platero, Phys. Rev. B 76, 073311 (2007); ibid. 78, 193310 (2008).

\bibitem{grid38} A. D. Chepelianskii et al., Eur.
Phys. J. B 60, 225 (2007).

\bibitem{grid39} A. Auerbach et al., Phys. Rev. B 76, 205318 (2007).

\bibitem{grid40} I. A. Dmitriev et al., Phys. Rev. B 75,
245320 (2007).

\bibitem{grid42} P. H. Rivera et al., Phys. Rev. B 79, 205406 (2009).

\bibitem{grid43} I. G. Finkler et al., Phys. Rev. B 79, 085315 (2009).

\bibitem{grid44} X. L. Lei et al., Appl. Phys. Lett. 94,
232107 (2009).

\bibitem{grid45} A. D. Chepelianskii et al., Phys. Rev. B 80,
241308(R) (2009).

\bibitem{grid46} I. A. Dmitriev et al., Phys. Rev. B 80, 165327 (2005).
\bibitem{grid47} T. Toyoda, Mod. Phys. Lett. B 24, 1923 (2010).
\bibitem{grid48} D. Hagenmuller et al.,
Phys. Rev. B 81, 235303 (2010).
\bibitem{grid49} J. Inarrea et al., Phys. Rev. B 82, 205321
(2010).

\end{thebibliography}

\newpage
\pagebreak
\pagebreak
\pagebreak





\end{document}